\documentclass[traditabstract,rnote]{aa} 
\usepackage{graphicx}
\usepackage{natbib}
\bibpunct{(}{)}{;}{a}{}{,} 
\begin{document}
\title{Stellar Limb-Darkening Coefficients for CoRot and Kepler}

\author{D. K. Sing\inst{1,2}}

   \offprints{D. K. Sing}

\institute{
Institut d'Astrophysique de Paris, CNRS/UPMC, 98bis
  boulevard Arago, 75014 Paris, France
\and
 Astrophysics Group, School of Physics, University of Exeter, Stocker
Road, Exeter EX4 4QL
\\
}

   \date{Received 16 November 2009 / Accepted 11 December 2009}

  \abstract{
Transiting exoplanets provide unparalleled access to the
fundamental parameters of both extrasolar planets and their host
stars.  We present limb-darkening coefficients (LDCs) for the
exoplanet hunting CoRot and Kepler missions.  The LDCs are calculated with ATLAS stellar
atmospheric model grids and span a wide range of
T$_{\rm eff}$, log g, and metallically [M/H].   Both CoRot and Kepler contain
wide, nonstandard response functions, and are producing a large
inventory of high-quality transiting lightcurves, sensitive to stellar
limb darkening.
Comparing the stellar model limb darkening to results from the first seven CoRot planets, we find better fits 
are found when two model intensities at the limb are excluded in the
coefficient calculations.  This calculation method can help to 
avoid a major deficiency present at the limbs of the 1D stellar models.
}

\titlerunning{Stellar Limb-Darkening Coefficients for CoRot \& Kepler}
\authorrunning{Sing}
   \maketitle

%

\section{Introduction}
Transiting exoplanets have provided an unprecedented opportunity to
directly measure physical parameters of the planet and host star, such
as mass and radius.  The lightcurve shape of a transit (where the planet passes
in front of the star as viewed from the Earth) is primarily determined
by the planet-to-star radius ratio, impact parameter, and stellar
limb darkening.  Thus, an accurate treatment of limb darkening is
critical when deriving planetary radii and transmission spectra from
transit data.

With high signal-to-noise (S/N) transit light curves, the
limb darkening can be fit and compared to theoretical stellar
atmospheric models, providing a method to scrutinize and test different
models (e.g. \citealt{2008MNRAS.386.1644S}).  
In a comparison with the sun, \cite{2008ApJ...686..658S} found that for the widely used Kurucz 1D
ATLAS stellar models\footnote{http://kurucz.harvard.edu/}, the largest differences between existing
limb-darkening data \citep{1994SoPh..153...91N} and the 1D stellar models was at the limb,
where ATLAS models predict a dramatic increase in the strength of limb
darkening.  For the sun, the ATLAS models over-predict the strength of
limb darkening by $>$20\% at $\mu=cos(\theta)$ values below 0.05,
though seemed to perform well otherwise, over-predicting the strength by
only a few percent.
While there remain few other observational constraints,
  theoretical atmospheric models have also been unable to provide a
  satisfactory fits to the 
observations of eclipsing binary stars \citep{2008A&A...482..259C} and the
transiting data of HD209458 \citep{2009A&A...506.1335C}.  However, different LDC
calculation methods may help improve the situation, with
\citep{2009A&A...505..891S} recently fitting near-infrared LDCs to a
level of a few percent with HST/NICMOS transit data of HD 189733, finding good
agreement with the theoretical stellar models.

The exoplanet hunting missions CoRot and Kepler provide an excellent opportunity
to further test these stellar models, as high S/N transits for a variety of spectral
types are being discovered.  Both CoRot and Kepler operate at optical
wavelengths, where the effects of limb-darkening are strong, using
wide band filters with wavelengths between $\sim$4000 and $\sim$9000
\AA.  Here we present stellar limb-darkening coefficients for both the CoRot
and Kepler satellites, along with some initial results comparing 
different calculation methods to the observed
stellar LD.  We present our calculation methods in Sect. 2, compare
our results to CoRot transit data in Sect. 3, and make concluding
remarks in Sect. 4.

\section{The numerical methods}

We calculate LDCs for the laws most commonly used in exoplanetary transit
work:\\
The Linear law
\begin{equation}  \frac{I(\mu)}{I(1)}=1 - u(1 - \mu).\end{equation}
The Quadratic law
\begin{equation}  \frac{I(\mu)}{I(1)}=1 - a(1 - \mu) - b(1 - \mu)^2 .\end{equation}
The Non-Linear law
\begin{equation}  \frac{I(\mu)}{I(1)}=1 - c_1(1 - \mu^{1/2}) - c_2(1 - \mu) - c_3(1 - \mu^{3/2}) - c_4(1 - \mu^{2}), \end{equation}
where $I(1)$ is the intensity at the center of the stellar disk,
$\mu=cos(\theta)$ (which is the angle between the line of sight and the
emergent intensity), while $u$, $a$, $b$, and $c_n$ are the LDCs.  These laws can all
be used along with the analytical transit light models of
\cite{2002ApJ...580L.171M}, and the definitions of each law have been set
to conform with the work of \cite{2000A&A...363.1081C}.

In addition to the above laws, we also calculate a variant of Equation
(3), which is a three parameter non-linear law, 
\begin{equation}  \frac{I(\mu)}{I(1)}=1 - c_2(1 - \mu) - c_3(1 - \mu^{3/2}) - c_4(1 - \mu^{2}), \end{equation}
introduced by \cite{2009A&A...505..891S} to improve the performance
of the calculated LD at the limb, while still providing enough flexibility to
capture the inherently non-linear nature of stellar LD.
The $\mu^{1/2}$ term from the four parameter non-linear law mainly
affects the intensity distribution at small $\mu$ values and is not
needed when the intensity at the limb varies approximately linearly at small $\mu$ values.  
Dropping the $c_1$ term is also further motivated by both solar data \citep{1994SoPh..153...91N} and 3D stellar models
\citep{2006A&A...446..635B}, which show the intensity distribution at the
limb to vary smoothly to $\mu=0$, with no dramatic or sudden increases
in limb-darkening strength as observed in the ATLAS models.  

We choose a least squares method to fit for the limb-darkening
coefficients from ATLAS models.
The model specific intensities were first integrated at each angle
using the CoRot PF white-light \citep{2009A&A...506..411A} and
Kepler\footnote{http://keplergo.arc.nasa.gov/CalibrationResponse.shtml}
response functions.
We used only values of $\mu \geq 0.05$
for the linear, quadratic, and three parameter non-linear laws
(corresponding to 15 angles in the ATLAS model grids) while retaining all 17 angles,
including the limb intensities, for the four-parameter non-linear law.

The four-parameter law best describes the tabulated ATLAS model grid intensities 
and should be analogous to the widely used LDCs of \cite{2000A&A...363.1081C}. 
Thus in our study, the four-parameter non-linear law is the best
representation of the original stellar atmospheric models themselves, while the
linear, quadratic and three-parameter non-linear laws are calculated with the
intent of improving the limb intensities.  The results of the
calculations are given in Tables 1 and 2, with the full versions of
these table available\footnote{see electronic version or http://vega.lpl.arizona.edu/$\sim$singd/}.
In Fig. 1, we illustrate typical CoRot and Kepler model limb darkening profiles for F, G,
and K main sequence stars.

\begin{figure}
 {\centering
  \includegraphics[width=0.35\textwidth,angle=90]{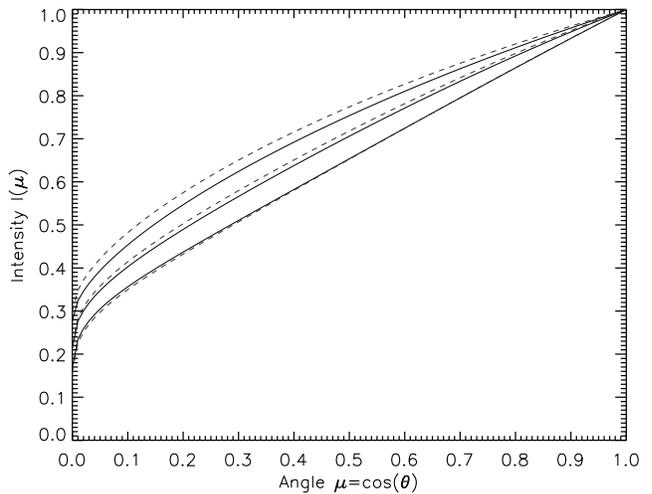}}
\caption[]{Stellar limb darkening for CoRot (solid lines) and Kepler (dashed
  lines) calculated from Atlas models appropriate for a solar metallically F5V,
  G5V, and K5V star (top to bottom: T$_{\rm eff}$=6500, 5500, 4500 K; log
  g=4.5, [M/H]=0.0).  The Atlas models show earlier type stars
  exhibiting increasingly non-linear intensity distributions for
  the CoRot and Kepler bandpasses.
}
\end{figure}


\section{Comparisons with CoRot exoplanet transit results}

We compared the calculated LDCs with results from six of the first seven
CoRot exoplanets, four of which have LDCs determined from the transit
light curve fit: CoRot-1 \citep{2008A&A...482L..17B}, CoRot-2
\citep{2008A&A...482L..21A}, CoRot-3 \citep{2008A&A...491..889D},
and CoRot-4 \citep{2008A&A...488L..43A}.  The LDCs of CoRot-5 and
CoRot-7 were not fit by the authors \citep{2009A&A...506..287L,
  2009A&A...506..281R}, so no comparison was possible.  With the limb intensities a
potential source of disagreement between the observations and models, we compared the
theoretical values of I($\mu$=$0$) with the CoRot results (see
Fig. 2). 
This comparison is aided by the fact that for each law considered here,
$I(0)$=$1-\Sigma$($C_n$), making a direct comparison between
different laws possible.  In addition, the
authors of CoRot 1, 2, \& 3 choose to fit the
LDC $u_+=a+b$, for the quadratic law, which is directly related to the
limb intensity, $I(0)$=$1-u_{+}$.  CoRot-4 was fit with the linear law.
We also calculated the model uncertainty for the theoretical LDCs,
incorporating the quoted uncertainty in stellar parameters (T$_{\rm eff}$, log g,
  [M/H]) using the partial derivative of the coefficients as a function
  of the stellar parameters.  We find that typical errors in T$_{\rm eff}$, log g,
  and [M/H] lead to only a small change in model LDCs for the linear
  and quadratic laws, with the model limb
  intensities uncertain by only a few percent or less.  This error
  estimation breaks down for the
  higher-order laws, as fitting degeneracies between the coefficients
  lead to unrealistically large partial derivatives and large errors.
  In those cases, we used the uncertainty from the linear law
  as a reasonable estimate of the model $I(0)$ error.  However, the
  model error is small and much less important than the observational errors.
 
From Fig. 2, the most obvious disagreement between the models and
observations is with the active star CoRot-2 \citep{2008A&A...482L..21A}.  A determination of the
planet-to-star radius ratio is affected by stellar activity, as shown
for CoRot-2 by \cite{2009A&A...505.1277C} who re-determined the
planetary radius, taking into account stellar activity, finding a
larger radius than either \cite{2008A&A...482L..21A} or \cite{2009arXiv0911.5087G}.  However, as
the radii and limb darkening are linked in a transit fit, the limb
darkening coefficients are also affected by stellar activity, though these
parameters were not re-determined.  With these complications due to
stellar activity, a proper comparison of limb darkening will likely have
to wait until the limb darkening is also re-determined in conjunction
with the planetary radii.

The three CoRot targets (1, 3 \& 4) are sufficient to see a
significant improvement in model limb intensities, when calculating LDCs with
15-angles and using lower order laws.  For every CoRot target thus
far, the calculated limb intensities of the ATLAS models are
significantly lower than the observed transit fit values, further
proof of the model limb deficiencies previously mentioned.

\begin{figure*}
 {\centering
  \includegraphics[width=0.58\textwidth,angle=90]{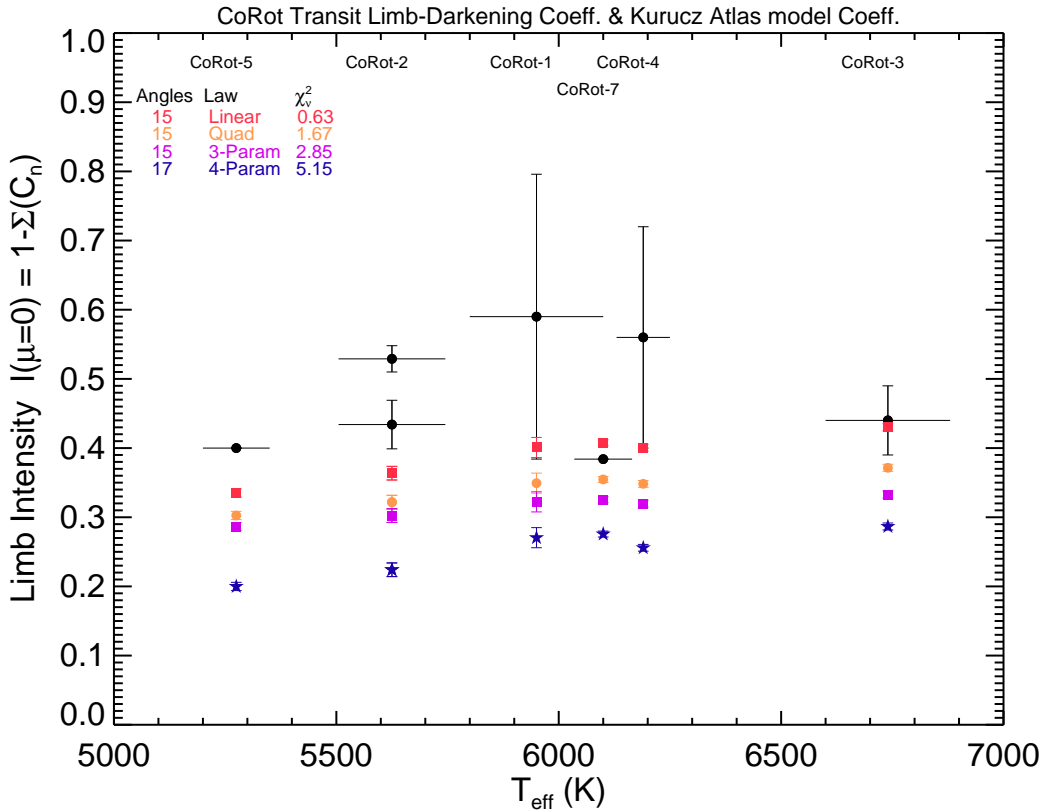}}
\caption[]{Limb intensity for the CoRot transiting planets (black circles) vs.
  effective stellar temperature.  CoRot-5 and CoRot-7 did not have
  lightcurve-fit
  LDCs, and the adopted values without y-error bars are shown.  The
  ATLAS model limb intensities for the linear (red), quadratic (orange),
  three-parameter non-linear (purple), and four-parameter non-linear
  (blue stars) laws are also shown, along with the error derived from the
  uncertainty in the stellar parameters.  The model errors are
  small and typically less than the size of the plotted point.  Our
  method of calculating for LDCs, with lower order laws and excluding the model limb intensities,
  shows a better agreement with the observations.
}
\end{figure*}

There is a potential source of disagreement between comparing $I(0) $ when
different laws are used in the stellar model LD and transit-fit LD.
In our model calculations, the general trend is to decrease $I(0)$ when
adopting a higher order law to describe the intensity.
Higher order laws are increasingly sensitive to the shape of the falling intensity
at the limb, thus reducing the fit value of $I(0)$.
However, we believe this to be a second order effect which should not seriously
affect these results, especially as several of the CoRot LDCs are
poorly constrained and are not sensitive to higher order terms. 

Comparing the $I(0)$ calculated from the different laws to the fit CoRot values, the 17-angle four parameter non-linear
law has a $\chi^2_{\nu}$ of 5.15, the 15-angle 3-parameter law has a 
$\chi^2_{\nu}$ of 2.85, the quadratic law a $\chi^2_{\nu}$ of 1.67 and the linear
law a $\chi^2_{\nu}$ of 0.63.  For the lower order laws, a better fit
to $I(0)$ is obtained by using 15-angles to calculate the LDCs, disregarding the
model limb values.  Though the linear law represents the best fit for $I(0)$ here,
the expected deviations of the model $I(\mu)$ from a linear trend in
F and G type stars (Fig. 1) will likely make higher order laws
necessary to describe the full intensity distribution.

Determining which limb darkening law to use in a transit light curve
fit is very dependent on the S/N of the data and the
stellar type of the star being studied.  From our model calculations,
we expect that linear laws will likely fit transit light curves quite
well for both CoRot and Kepler in cooler K type stars, or for earlier
types at low S/N.  Higher order terms will likely be needed in F and G
type stars at high S/N, where multiple LDCs can be fit at good
precision.   

\begin{table*} 
\caption{CoRot Stellar Limb-Darkening Coefficients}
\label{table:1}
\begin{centering}
\renewcommand{\footnoterule}{}  
\begin{tabular}{lllllllllllll} 
\hline\hline  
T$_{\rm eff}$  & Logg  & [M/H]& linear & \multicolumn{2}{c}{\underline{~~~~quadratic~~~~}}   &  \multicolumn{3}{c}{\underline{~~3 parameter non-linear~~} }&    \multicolumn{4}{c}{\underline{~~~~~~~~~4 parameter non-linear~~~~~~~~~}} \\ 
(K)       &         &         & $u$   & $a$ & $b$             &$c_2$ & $c_3$& $c_4$   &  $c_1$  &  $c_2$  &  $c_3$   & $c_4$ \\
\hline 
 4000  &   4.50 &    0.00 &  0.6690 &  0.9361 & -0.2241 &  1.6776 & -1.5366 &  0.6067 &  0.5651 &  0.1233 &  0.2108 & -0.0817 \\
 4250  &   4.50 &    0.00 &  0.7091 &  0.8210 & -0.0939 &  1.2813 & -0.9538 &  0.4217 &  0.6895 & -0.6083 &  1.1653 & -0.4118 \\
 4500  &   4.50 &    0.00 &  0.7104 &  0.7955 & -0.0714 &  1.1266 & -0.6859 &  0.2995 &  0.7336 & -0.8830 &  1.5672 & -0.5866 \\
 4750  &   4.50 &    0.00 &  0.6975 &  0.8143 & -0.0980 &  1.1159 & -0.6250 &  0.2399 &  0.7099 & -0.8275 &  1.5532 & -0.6164 \\
 5000  &   4.50 &    0.00 &  0.6831 &  0.8401 & -0.1317 &  1.1434 & -0.6285 &  0.2081 &  0.6995 & -0.7701 &  1.5147 & -0.6341 \\
 5250  &   4.50 &    0.00 &  0.6656 &  0.8654 & -0.1677 &  1.1917 & -0.6760 &  0.1978 &  0.6904 & -0.6941 &  1.4338 & -0.6306 \\
 5500  &   4.50 &    0.00 &  0.6464 &  0.8875 & -0.2023 &  1.2571 & -0.7658 &  0.2118 &  0.6560 & -0.5308 &  1.2309 & -0.5712 \\
 5750  &   4.50 &    0.00 &  0.6280 &  0.9036 & -0.2312 &  1.3297 & -0.8831 &  0.2462 &  0.6060 & -0.3196 &  0.9569 & -0.4745 \\
 6000  &   4.50 &    0.00 &  0.6096 &  0.9177 & -0.2584 &  1.4350 & -1.0719 &  0.3211 &  0.5409 & -0.0366 &  0.5688 & -0.3213 \\
 6250  &   4.50 &    0.00 &  0.5937 &  0.9244 & -0.2774 &  1.5321 & -1.2593 &  0.4034 &  0.4857 &  0.2088 &  0.2177 & -0.1753 \\
 6500  &   4.50 &    0.00 &  0.5797 &  0.9300 & -0.2939 &  1.6369 & -1.4647 &  0.4980 &  0.4256 &  0.4731 & -0.1624 & -0.0132 \\
 6750  &   4.50 &    0.00 &  0.5687 &  0.9341 & -0.3065 &  1.7227 & -1.6343 &  0.5770 &  0.3898 &  0.6510 & -0.4303 &  0.1030 \\
 7000  &   4.50 &    0.00 &  0.5602 &  0.9394 & -0.3180 &  1.7944 & -1.7717 &  0.6398 &  0.3682 &  0.7769 & -0.6243 &  0.1868 \\
 7250  &   4.50 &    0.00 &  0.5531 &  0.9463 & -0.3299 &  1.8515 & -1.8757 &  0.6842 &  0.3622 &  0.8469 & -0.7397 &  0.2348 \\
 7500  &   4.50 &    0.00 &  0.5479 &  0.9566 & -0.3429 &  1.8972 & -1.9489 &  0.7108 &  0.3552 &  0.9092 & -0.8296 &  0.2673 \\
 \hline
\end{tabular}
\end{centering}
\\
Note:  The full version of this table can be found in the electronic version online
and at http://vega.lpl.arizona.edu/$\sim$singd/ which also includes an IDL
program to read and interpolate the table.   The full table
covers: T$_{\rm eff}$ from 3500 to 50000 K, log g from 0.0 to 5.0, and log
[M/H] from +1 to -0.5 with a turbulent velocity of 2 km s$^{-1}$.  The
linear, quadratic, and 3 parameter non-linear laws are calculated
with the improved calculation method using 15-ATLAS angles with the limb intensities excluded, while the 4 parameter non-linear law was calculated
with all 17 angles including those at the limb (see text).
\\
\caption{Kepler Stellar Limb-Darkening Coefficients}
\label{table:2}
\begin{centering}
\renewcommand{\footnoterule}{}  
\begin{tabular}{lllllllllllll} 
\hline\hline  
T$_{\rm eff}$ & Logg  & [M/H]& linear & \multicolumn{2}{c}{\underline{~~~~~quadratic~~~~~}}   &  \multicolumn{3}{c}{\underline{~~~3 parameter non-linear~~~} }&    \multicolumn{4}{c}{\underline{~~~~~~~~~4 parameter non-linear~~~~~~~~~}} \\ 
(K)       &         &         & $u$   & $a$ & $b$             &$c_2$ & $c_3$& $c_4$   &  $c_1$  &  $c_2$  &  $c_3$   & $c_4$ \\
\hline 
 4000  &   4.50  &   0.00 &  0.6888 &  0.5079 &  0.2239 &  1.6669 & -1.4738 &  0.5729 &  0.5478 &  0.1602 &  0.2201 & -0.0944 \\
 4250  &   4.50  &   0.00 &  0.7215 &  0.6408 &  0.0999 &  1.2877 & -0.9263 &  0.4009 &  0.6928 & -0.6109 &  1.2029 & -0.4366 \\
 4500  &   4.50  &   0.00 &  0.7163 &  0.6483 &  0.0842 &  1.1412 & -0.6724 &  0.2793 &  0.7393 & -0.8838 &  1.5980 & -0.6136 \\
 4750  &   4.50  &   0.00 &  0.6977 &  0.6036 &  0.1164 &  1.1431 & -0.6355 &  0.2272 &  0.7156 & -0.8156 &  1.5595 & -0.6357 \\
 5000  &   4.50  &   0.00 &  0.6779 &  0.5528 &  0.1548 &  1.1854 & -0.6693 &  0.2070 &  0.7072 & -0.7489 &  1.4971 & -0.6442 \\
 5250  &   4.50  &   0.00 &  0.6550 &  0.4984 &  0.1939 &  1.2500 & -0.7540 &  0.2138 &  0.7011 & -0.6649 &  1.3885 & -0.6275 \\
 5500  &   4.50  &   0.00 &  0.6307 &  0.4451 &  0.2297 &  1.3274 & -0.8765 &  0.2442 &  0.6716 & -0.5029 &  1.1678 & -0.5574 \\
 5750  &   4.50  &   0.00 &  0.6074 &  0.3985 &  0.2586 &  1.4038 & -1.0114 &  0.2882 &  0.6283 & -0.3063 &  0.8965 & -0.4593 \\
 6000  &   4.50  &   0.00 &  0.5842 &  0.3539 &  0.2851 &  1.5101 & -1.2143 &  0.3714 &  0.5685 & -0.0364 &  0.5100 & -0.3038 \\
 6250  &   4.50  &   0.00 &  0.5640 &  0.3198 &  0.3023 &  1.6062 & -1.4130 &  0.4616 &  0.5170 &  0.1979 &  0.1587 & -0.1541 \\
 6500  &   4.50  &   0.00 &  0.5459 &  0.2901 &  0.3167 &  1.7070 & -1.6236 &  0.5611 &  0.4576 &  0.4564 & -0.2249 &  0.0122 \\
 6750  &   4.50  &   0.00 &  0.5312 &  0.2672 &  0.3267 &  1.7862 & -1.7936 &  0.6430 &  0.4219 &  0.6279 & -0.4938 &  0.1317 \\
 7000  &   4.50  &   0.00 &  0.5191 &  0.2478 &  0.3358 &  1.8587 & -1.9464 &  0.7165 &  0.3968 &  0.7640 & -0.7134 &  0.2302 \\
 7250  &   4.50  &   0.00 &  0.5085 &  0.2308 &  0.3437 &  1.9144 & -2.0641 &  0.7722 &  0.3905 &  0.8337 & -0.8442 &  0.2902 \\
 7500  &   4.50  &   0.00 &  0.5003 &  0.2165 &  0.3512 &  1.9533 & -2.1433 &  0.8075 &  0.3875 &  0.8777 & -0.9264 &  0.3259 \\
 \hline
 \end{tabular}
\end{centering}
\\
Note:  Same as note for Table 1.  
\end{table*}


\section{Conclusions}

For use within the community, we present ATLAS model grid LDCs
calculated for the CoRot and Kepler transiting exoplanet missions.
We find better agreement between the existing CoRot
observations and model LDC when two limb intensities are not used in
the calculations, and incorporate this method in the
presented LDCs.

The future catalog of transiting planets discovered by CoRot and
Kepler offers the
prospect of substantially improving the theoretical models of stellar
limb darkening.  The very high photometric precision of Kepler should
allow for multiple LDCs to be fit at the percent level, which should open up
many detailed comparisons with stellar atmospheric models.  For
instance, Kepler (at high temporal resolution) will be quite sensitive to non-linear LD
terms, as there will be sufficient S/N to accurately fit for coefficients
beyond just the linear term.  The LDCs presented here are intended to
aid in these studies and be of general use to the community.


\begin{acknowledgements}
D.K.S. is supported by CNES.  We thank Robert Kurucz for making his
grid of stellar models publicly available and the referee for their
helpful comments.

\end{acknowledgements}


\bibliographystyle{aa} 
\bibliography{Sing} 

\end{document}